\documentclass{PoS}

\def\Journal#1#2#3#4{{\it #1} {\bf #2}, #3 (#4)}

\title{Diffractive Higgs production}

\ShortTitle{Diffractive Higgs}

\author{\speaker{Jeffrey R. Forshaw}\\
        School of Physics \& Astronomy \\ University of Manchester, Manchester M13 9PL, UK\\
        E-mail: \email{forshaw@mail.cern.ch}}

\abstract{
Instrumenting the LHC to measure the outgoing protons in the process $pp \to p+X+p$ opens
up new possibilities for physics studies in the Standard Model and beyond. In this talk I first present
an overview of the underlying QCD calculation and its uncertainties. I then move on to present a variety of 
new physics scenarios which one would want to explore via this unique channel. There is a short discussion
of the recent CDF data which may provide evidence for the existence of central exclusive jet production
at the Tevatron.}

\FullConference{International Workshop on Diffraction in High-Energy Physics -DIFFRACTION 2006 -\\
		 September 5-10 2006\\
		 Adamantas, Milos island, Greece}

\begin{document}

\section{Introduction}
The central exclusive production (CEP) of a system $X$ in the reaction $pp \to p+X+p$ is of growing 
theoretical \cite{JRF} and experimental interest \cite{Cox,Albrow,Croft}. 
There is a real possibility to install low angle proton detectors 
at the LHC in order to measure the final state protons and hence the invariant mass of the central 
system $X$. It is hoped that this mass $M_X$ may be measured to a precision approaching 1 GeV. Most 
interesting is to exploit this channel in order to detect processes which may be very difficult or 
even impossible to measure using more conventional channels. In the following sections I will discuss 
a sample of potentially interesting scenarios in which the system $X$ could be a single Higgs boson 
(Standard Model or supersymmetric) or a pair of gluinos. 

\begin{figure}[h]
\begin{center}
\includegraphics[width=2.0in]{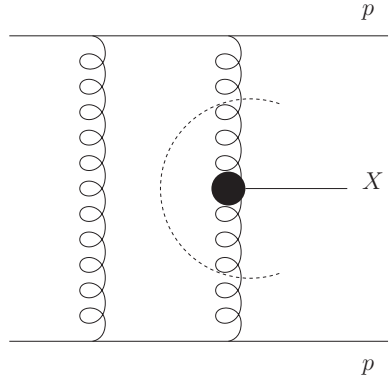}
\caption{Central exclusive production of a system $X$. The dotted line illustrates the factorization of
the production of $X$ from the gluon luminosity.}
\label{fig:CEP}
\end{center}
\end{figure}

\begin{figure}[h]
\begin{center}
\includegraphics[width=4.0in]{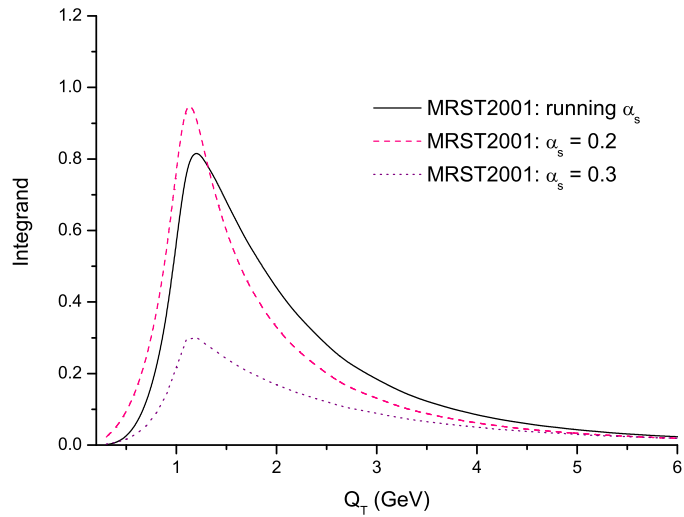}
\caption{The integrand of the $Q$ integral for the production of a 100 GeV SM Higgs boson.}
\label{fig:integrand}
\end{center}
\end{figure}

Apart from the challenges related to building and installing low angle proton detectors in the LHC,
the theorists are charged with developing the requisite theory. For a recent and more detailed 
review than is provided here see \cite{JRF}. The cross-section factorises
into a hard scattering part which represents the gluon fusion sub-process $gg \to X$,
$d\hat{\sigma}(y,\hat{s})$, and an effective gluon luminosity, $d\mathcal{L}
(\hat{s},y)/(dy~d\hat{s})$, as illustrated in Fig.\ref{fig:CEP}. 
As a result, we can write the cross-section for
producing any central system $X$ at rapidity $y$ and invariant mass $\hat{s}$ as
\begin{equation}
\frac{d\sigma}{dyd\hat{s}}=\frac{d\mathcal{L}(\hat{s},y)}{dyd\hat{s}}
d\hat{\sigma}(y,\hat{s})\;. \label{eq:exclusive}
\end{equation}
where
\begin{equation}
\hat{s}\frac{d\mathcal{L}(\hat{s},y)}{dyd\hat{s}}=\left[  \frac{\pi}{8b}
{\displaystyle\int\limits^{M^{2}/4}}
\frac{dQ^{2}}{Q^{4}}f(x_{1},Q,M)f(x_{2},Q,M)\right]  ^{2} \label{eq:lumi}
\end{equation}
and
\begin{equation}
f(x,Q,M)=R_{g}Q^{2}\frac{\partial}{\partial Q^{2}}\left(  \sqrt{T(Q,M)}
xg(x,Q^{2})\right)
\end{equation}
with the Sudakov factor
\begin{equation}
T(Q,M)=\exp\left(  -\int_{Q_{T}^{2}}^{M^{2}/4}\frac{dp_{T}^{2}}{p_{T}^{2}
}\frac{\alpha_{s}(p_{T}^{2})}{2\pi}\int_{0}^{(1+2p_{T}/M)^{-1}}dz\;[zP_{gg}
(z)+\sum_{q}P_{qg}(z)]\right)~.
\end{equation}
In this expression we have assumed a dependence on the $p_T$ of the scattered proton of
$\exp(-bp_T^2)$ and $g(x,Q^{2})$ is the gluon distribution function of the proton. 
$R_{g}$ is a parameter which corrects for the fact that we really need an off-diagonal 
gluon distribution. For SM Higgs production at the LHC one expects 
$R_{g}\approx1.2$ \cite{KMR}. Formally the $Q$
integral needs to be cut-off in order to avoid the pole in the running
coupling. In practice the integral is peaked well above $\Lambda_{\mathrm{QCD}}$,
as illustrated in Fig.\ref{fig:integrand}, 
and so the final results are insensitive to the value of the infra-red cutoff.
In the limit that the scattered protons leave at zero angle it is noteworthy that 
the fusing gluons have equal helicities and hence the central system is predominantly produced 
with $J_z=0$ and even parity. This selection rule is advantageous in suppressing background 
and it can be used to learn about the CP structure of $X$. The principal theoretical uncertainty 
resides in the calculation of the luminosity function. 

\begin{figure}[h]
\begin{center}
\includegraphics[width=4.0in]{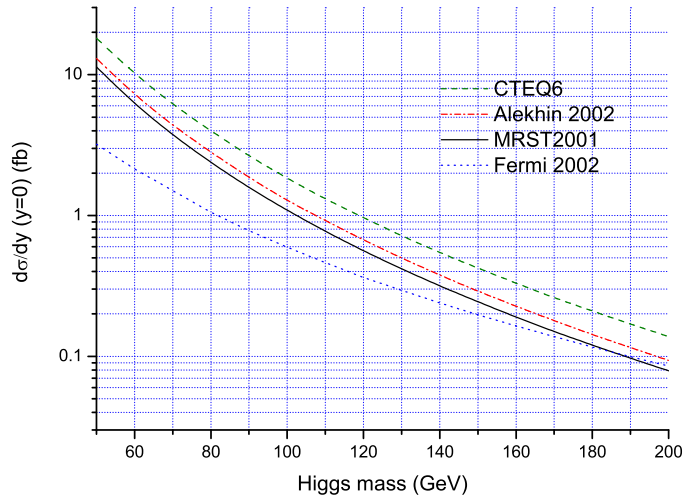}
\caption{The cross-section dependence on the choice of PDF for the production of a 100 GeV SM Higgs boson.}
\label{fig:pdfs}
\end{center}
\end{figure}

Except for the so-called ``gap survival'' factor (of which more a little later), Fig.\ref{fig:integrand} 
shows that the calculation can be performed using perturbation 
theory (the plot is for a 100 GeV SM Higgs). The mass of the central system squeezes the dynamics to 
short distances, although not to distances $\sim 1/M_X$ since the screening gluon (the passive gluon
in Fig.\ref{fig:CEP}) has the effect of pushing the physics into the infra-red. 
The infra-red physics would win if it were not for the fact that we
must also insist that there be no radiation off the exchanged gluons. Including this non-emission
requirement introduces the Sudakov factor $T(Q,M)$ above. This has the effect of taming the infra-red,
with the result that for a $\sim 100$ GeV central
system the typical virtuality flowing around the gluon loop is $Q \sim 2$ GeV. Technically, the 
non-emission is accounted for by summing logarithmic terms like $\sim (\alpha_s \ln^2(M_X^2/Q^2))^n$ 
to all orders $n$. In fact one must go beyond this ``double log'' accuracy and sum also single logarithms.
In so doing one gains more than an order of magnitude enhancement in the final cross-section.
In addition, one also needs knowledge of the two-gluon coupling into the proton (a ``skewed 
parton distribution function''). Small uncertainties here are magnified in the cross-section since they 
appear in the fourth power. Studies indicate something like a factor two uncertainty from this source 
\cite{JRF} although it may be more \cite{Leif}, see Fig.\ref{fig:pdfs}.

\begin{figure}[h]
\begin{center}
\includegraphics[width=1.5in]{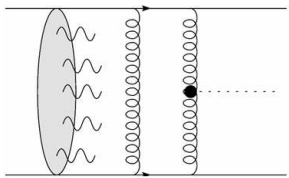}
\caption{Illustration of a soft rescattering which could fill in any rapidity gap.}
\label{fig:sue}
\end{center}
\end{figure}

Finally, we must address ``gap survival''. Apart from the short distance physics we have just been
discussing, there is also the chance that the external protons can undergo some soft re-scattering which
will produce additional particles, as illustrated in Fig.\ref{fig:sue}. 
The probability of that not happening is presumably well approximated
by assuming it to be independent of the short distance dynamics. In which case it would factorize into
an overall re-scaling of the final cross-section: this re-scaling is called the ``gap survival factor''.
Its value is not something we can compute reliably. Within a particular model one can try to use
data on elastic scattering, total cross-sections and inelastic scattering in order to estimate the
effects of soft re-scattering. Eikonal models are of this type and they have also met with 
some success in explaining the features of the soft underlying event in hadronic scattering 
processes \cite{Borozan,Sjostrand,Odagiri}.
The value of the survival factor can also be estimated by comparing the rates for diffraction at HERA
with corresponding rates at the Tevatron. This has been done, in particular for diffractive dijet
production \cite{CDF:dijets}, and the lower rate at the Tevatron is consistent with eikonal model 
expectations \cite{Kaidalov,Maor}.
Fortunately, one can try and extract the survival factor from data itself and thereby minmize the
uncertainty in theoretical predictions of CEP, e.g. one could use $pp \to p+\gamma \gamma+ p$ to
callibrate other CEP processes.

I have been describing the ideas which underpin the theoretical calculations pioneered by the 
Durham group \cite{KMR}. These calculations have been faithfully implemented in the ExHuME Monte 
Carlo event generator \cite{Exhume}. There are other approaches in the literature which 
have also been implemented in event generators: the EDDE code implements the calculations of 
Ryutin \& Petrov \cite{EDDE} whilst the DPEMC 
code implements the calculations of the Saclay group \cite{DPEMC}.

\section{Higgs bosons}
Perhaps the most interesting possibility is if $X$ is a single Higgs boson in which case the
calculation is precisely as presented in the introduction. Of course one would
hope to discover and measure the Higgs through inclusive production where the rate is much higher.
Nevertheless, CEP could provide important additional information on the mass, spin and parity
of the Higgs. Moreoever, if the Higgs is non-standard in some way then CEP may prove crucial in
understanding its nature; as we shall shortly see.
\subsection{Standard Model}
Remarkably, it is possible to observe the Higgs through its decay to $b$-quarks. Normally the QCD background
would be prohibitive however the QCD $b\bar{b}$ background is heavily suppressed by the spin selection
rule (the rate would in fact vanish for massless quarks) and one can aim to identify a clean enough
sample of $H \to b\bar{b}$. Studies indicate that a $S/B>1$ could be obtained with something like
10 signal events from 30 fb$^{-1}$ of data \cite{KMR}. 
Note that if one could measure CEP when the LHC is running at
high luminosity then this rate would increase very significantly. The main challenge is not the rate,
it is in triggering. Proton detectors at 420m cannot easily be included in the level 1 trigger and one
is thus forced to look for ways to trigger using the central detector.

One can avoid problems with the trigger by looking at the decay $H\to WW$. The rate becomes large
enough for masses above 120 GeV even at low luminosity running and studies indicate that the
backgrounds should be under control \cite{WW}. Note that for a clean enough sample, one can measure the Higgs
mass on an event-by-event basis. It is therefore not necessary to collect many events before a
competetive mass measurement is possible.

\subsection{Intense coupling regime of the MSSM}
To showcase the potential virtues of CEP, one does not need to hunt for obscure regions of new physics
parameter space. The intense coupling regime of the MSSM \cite{Boos} is a difficult region for studies via
inclusive channels since the three Higgs bosons are almost degenerate in mass and the largeness of
$\tan \beta$ means that the branching ratio of the Higgs to $b$-quarks is enhanced. Providing one
can trigger on the decaying Higgs, the rate for CEP of such a non-standard Higgs is greatly
enhanced over the corresponding rates in the Standard Model. For example, with all three Higgses
around 130 GeV one expects 125 $H^0$ and 70 $h^0$ events with a background of $\sim 2$ events from
only 30 fb$^{-1}$ of data \cite{KKMR}. Moreoever, the spin selection rule suppresses the rate for $A^0$ 
production so one can be sure to have filtered out the CP even states.     

\subsection{CP violating MSSM}
Another large $\tan \beta$ scenario of interest is that which may occur in the CP violating MSSM.
In this case, radiatively induced explicit CP violation mixes the three supersymmetric Higgses
into states of indefinite CP \cite{CPX}. Again it is common for them to be nearly degenerate in mass. In this
case one can hope to unravel the Higgs sector by exploiting the excellent mass resolution of CEP
in conjunction with the inherent CP even filter. The largeness of $\tan \beta$ again leads to a
very significant enhancement of the rate over that for a Standard Model Higgs, so rate should
not be a problem \cite{Pilaftsis}.

\section{Gluinos}
As our final example, let's turn to supersymmetric gluino production in scenarios where the
gluino is effectively a stable particle. Such a scenario is fashionable within the framework
of ``split supersymmetry'' models \cite{splitsusy}. The production rate for open gluino pair production is 
enhanced by a colour factor and by the fact that this is a strongly interacting process.
Consequently the rate can be sufficiently large even for gluino masses as large as 350 GeV
provided one can make measurements when the LHC is running at high luminosity \cite{Coughlin}. The 
stable gluinos look like delayed muons in the detectors (they form R-hadrons whose mass is
almost equal to the gluino mass and which interact weakly with the calorimeters) and are
thus relatively easy to trigger on and detect. Moreoever, the unique signature makes it
possible to suppress the backgrounds sufficiently. Of course, one would produce gluino
pairs at a prolific rate in inclusive production however CEP has the crucial advantage that
one is able to use CEP events to extract the gluino mass on an event-by-event basis. Studies
have shown that with 300 fb$^{-1}$ of data one could hope to measure the gluino mass to
better than 2.5 GeV precision up to masses of 350 GeV \cite{Coughlin}. If stable gluinos do exist then the
Tevatron ought to be able to discover them using Run II data if they are as light as 350 GeV.
CEP would then be the perfect place to make a precision measurement of the mass.    

\section{CDF data}
\begin{figure}[h]
\begin{center}
\includegraphics[width=3.5in]{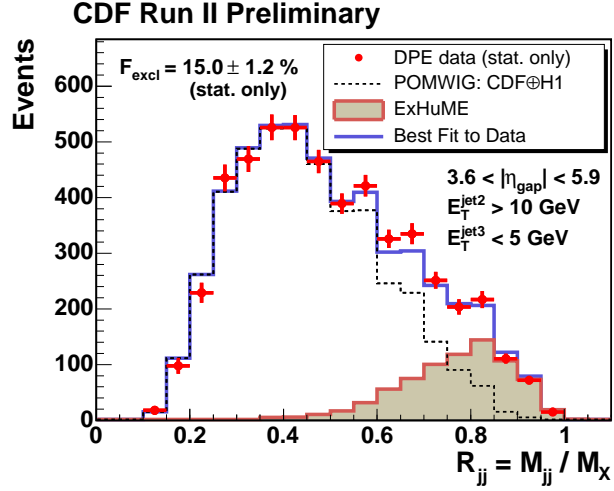}
\caption{The dijet mass fraction $R_{jj}$ measured by CDF and compared to theoretical expectations.}
\label{fig:rjj}
\end{center}
\end{figure}

\begin{figure}[h]
\begin{center}
\includegraphics[width=3.5in]{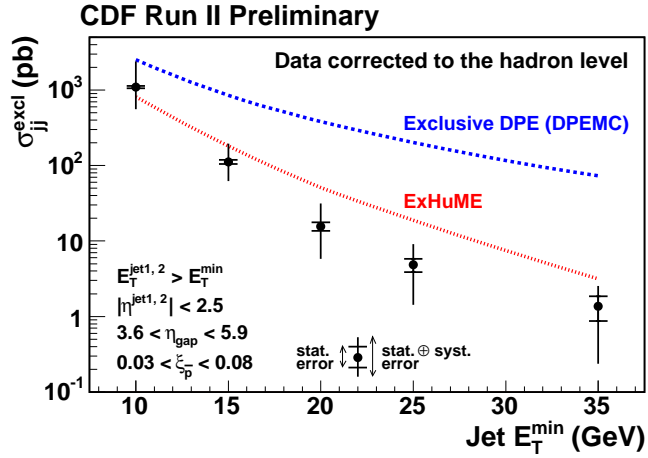}
\caption{The $E_{Tmin}$ distribution measured by CDF and compared to theoretical expectations.}
\label{fig:pt}
\end{center}
\end{figure}

In order to test our theoretical understanding it ought to be possible to search for CEP of dijet
events at the Tevatron. To that end, the results presented in the talk at this workshop by 
Goulianos are most interesting \cite{Dino}. The goal is to look for a sample of dijet events in
the reaction $pp \to p+jj+p$ in which almost all of the invariant mass of the central system is
contained in the dijets. In Fig.\ref{fig:rjj} the dijet mass fraction is plotted (with a veto on any
third jet), the Durham model calculation has been implemented in the ExHuME Monte Carlo \cite{Exhume} 
is shown as the shaded contribution in the figure. The dotted line shows the contribution from POMWIG,
an extension to HERWIG which simulates diffractive central inclusive processes of the type 
$pp \to p+jj+X+p$ \cite{POMWIG}, and there does appear to be
an excess of events at large $R_{jj}$ which can be explained by CEP. Shown in Fig.\ref{fig:pt} is the
cross-section for a pair of exclusively produced jets with $p_T > E_{Tmin}$. Again the ExHuME
Monte Carlo does a good job. Further evidence that CDF really is seeing CEP follows from the fact that
they also find a suppression of charm and bottom quark jets at 
high $R_{jj}$, which is in accord with expectations from CEP (it is a result of the spin selection rule).

\section{Conclusions}
Central exclusive production offers the possibility to explore new physics scenarios at the LHC
in a way which complements the mainstream LHC programme. It is worth mentioning that
there also exists a more conventional programme of physics which can be carried out using
tagged protons at the LHC. For example, one can use the protons as a source of real photons
and hence study $\gamma \gamma$ and $\gamma p$ collisions of known CM energy. Certainly
an exciting opportunity, occuring as it will many years before the $\gamma \gamma$ option
of a future linear collider.  Theoretical calculations are
challenging but can be performed using QCD perturbation theory and appear to be accurate to
within a factor $\sim 5$. Indeed, recent preliminary measurements by the CDF collaboration 
on central exclusive jet production do indicate the presence of exclusive dijets at a rate
consistent with theoretical expectations. We can surely look forward to a very exciting 
programme of research using low angle proton detectors at the LHC.   

\section*{Acknowledgments}
My sincere thanks are due to the organizers of a most interesting and enjoyable workshop.

\end{document}